\documentclass[%
 aip,
 amsmath,amssymb,
 reprint,%
]{revtex4-1}

\usepackage{graphicx}
\usepackage{dcolumn}
\usepackage{bm}
\usepackage[USenglish]{babel}
\usepackage[utf8]{inputenc}
\usepackage[T1]{fontenc}
\usepackage{mathptmx}
\usepackage{etoolbox}
\usepackage{xcolor}

\makeatletter
\def\@email#1#2{%
 \endgroup
 \patchcmd{\titleblock@produce}
  {\frontmatter@RRAPformat}
  {\frontmatter@RRAPformat{\produce@RRAP{*#1\href{mailto:#2}{#2}}}\frontmatter@RRAPformat}
  {}{}
}%
\makeatother
\begin{document}

\preprint{AIP/123-QED}

\title{Engineered Near- and Far-Field Optical Response of Dielectric Nanostructures using Focused Cylindrical Vector Beams}

\author{M. Montagnac}
\author{G. Agez}
\email{gonzague.agez@cemes.fr}
\author{A. Patoux}
\author{A. Arbouet}
\author{V. Paillard}
\email{vincent.paillard@cemes.fr}
\affiliation{CEMES-CNRS, Universit\'e de Toulouse, CNRS, UPS, Toulouse, France}

\date{\today}

\begin{abstract}
Near- and far-field optical properties of silicon nanostructures under linear polarization (Gaussian beam), and azimuthally or radially focused cylindrical vector beams are investigated by finite-difference time-domain method (FDTD) in Meep open-source software. A python toolkit allowing FDTD simulations in Meep for using those excitation sources is provided. In addition to the preferential excitation of specific electric or magnetic resonance modes as function of the excitation beam polarization, it is shown in the case of spheroids that shape anisotropy affects the resonance wavelength and the dipole orientation of the magnetic or electric dipole mode. For radial or linear polarization, the electric dipole resonance is split by an anapole mode depending on the spheroid symmetry axis with respect to the electric field orientation . Finally, the optical properties in both far-field (scattering pattern) and near-field (electric and magnetic field hot spots) can be tuned by changing the excitation polarization at a fixed wavelength and selecting properly the spheroid shape and dimensions. These numerical simulations can be extended to more complex shapes, or fabrication-friendly nanostructures such as nanocylinders with circular or elliptic sections.
\end{abstract}

\pacs{}

\maketitle


\section{Introduction} 

A plane wave can interact strongly with high refractive index dielectric particles of subwavelength dimension supporting optical resonances \cite{cao_tuning_2010, kuznetsov_optically_2016}. Because both electric and magnetic multipolar modes of comparable strengths can be excited in such nanostructures, several degrees of freedom can then be used to engineer the optical response of the nanoantenna \cite{kuznetsov_optically_2016, staude_tailoring_2013, liu_multipole_2020}, as well as its near-field coupling to quantum emitters  \cite{krasnok_spectroscopy_2018, bidault_dielectric_2019, poumirol_unveiling_2020}. Many works have been dedicated to improving the directivity of enhanced light scattering by studying the overlap between electric and magnetic multipole modes \cite{fu_directional_2013, wiecha_strongly_2017}, or creating a Huygens source out of dielectric nanoparticles \cite{bag_transverse_2018}.

The optical behavior is usually controlled by modifying the material \cite{baranov_all-dielectric_2017}, size  \cite{cao_tuning_2010,fu_directional_2013}, shape \cite{reena_tunable_2016, wiecha_strongly_2017}, and coupling \cite{yan_magnetically_2015, shibanuma_experimental_2017} of the nanoparticles, but alternative approaches involving focused cylindrical vector beams (CVB), such as higher-order Laguerre-Gaussian singular beams \cite{youngworth_focusing_2000, zhan_cylindrical_2009}, are being considered. Contrary to a linearly polarized excitation (plane wave or focused Gaussian beam), an azimuthally or radially polarized beam was used to selectively excite either magnetic or electric multipolar modes, leading to the  modification of the scattering spectrum and radiation pattern of a given individual nanoantenna \cite{banzer_experimental_2010, wozniak_selective_2015, manna_selective_2020, parker_excitation_2020}.

Recent studies deal with various dielectric or plasmonic nanostructures, from single nano-objects \cite{bautista_second-harmonic_2012, manna_selective_2017} to an assembly of a few nanoparticles \cite{yanai_near-_2014, reich_selection_2020, kroychuk_enhanced_2020}. In this paper, we present electrodynamical simulations applied to silicon nanospheres and nanospheroids (Si NS) to investigate the effect of shape anisotropy and the kind of CVB excitation on the behavior of the electric and magnetic dipole modes, as well as the anapole. We thus show that, by selecting the Si NS and CVB, it is possible to engineer the far-field properties, such as modifying the radiation pattern and scattering directivity or switching from a strongly radiating to a nonradiating mode at a fixed wavelength, as well as the near-field properties such as controlling, separating spatially, and enhancing electric and magnetic hotspots. 

The numerical simulations used for this present study were performed with the finite-difference time-domain (FDTD) method, using the open-source Meep software package \cite{oskooi_meep_2010}. We remind hereafter a brief description of two methods for generating radially and azimuthally polarized doughnut excitation in Meep ($i$) from Laguerre-Gaussian (LG) modes in the paraxial approximation, and ($ii$) from the calculation of a beam focused by a high-numerical-aperture (NA) aplanatic lens. We discuss their field of applications and provide  for both methods python script packages available in our Github repository  \cite{Agez_Git_2_2021}, allowing a straightforward implementation in Meep. Note that commercial software has already been used to perform FDTD simulations with CVB excitation \cite{manna_selective_2017, das_dark_2017} but, to the best of our knowledge, no open-source library was available for a direct utilization. We also point out that a purely analytical approach (generalized Lorenz-Mie theory) has been used in the case of high symmetry nanoparticles \cite{orlov_analytical_2012, zambrana-puyalto_role_2013, lukyanchuk_optimum_2015, parker_excitation_2020}. In this paper, the numerical approach was chosen in the perspective of investigating any kind of nanostructures, even of complex shape, that can be obtained through top-down fabrication techniques \cite{kuznetsov_optically_2016, wiecha_evolutionary_2017}.


\section{Focused cylindrical vector beams exciting a dielectric nanoparticle} 

\subsection{Focused cylindrical vector beam sources in Meep/FDTD}	
The cylindrical vector beams considered in the present study are characterized by both an azimuthally or radially distributed polarization and an intensity distribution having cylindrical symmetry. They can be seen as linear combination of a right-handed and a left-handed circularly polarized vortex beam with opposite orbital angular momentum ${l=\pm 1}$\cite{novotny_principles_2006}. Higher-order modes like Laguerre-Gaussian (LG) modes can be used to define an analytical expression in the paraxial regime and then be used in FDTD script to define the complex amplitude of the electric and/or magnetic source in a transverse plane before or after the waist position. From this initial condition, the beam propagation is calculated by solving the time-dependent Maxwell's equations.

However, for many applications, it is interesting to tightly focus a CVB in order to generate a strong longitudinal electric or magnetic field component and to spatially separate the electric and magnetic energy density \cite{kasperczyk_excitation_2015}. When the beam divergence angle becomes too large, by using a high numerical aperture (NA) objective for example, the fields computed using LG modes deviate from the expected behavior at the focus point and a more general (nonparaxial) approach has to be considered. For this purpose, several methods have been proposed based on $n$-order corrected expression of the field components \cite{cerjan_orbital_2011, yan_description_2007, barnett_orbital_1994} or using the scalar-complex-source model \cite{orlov2010complex}.

In practice, tightly focused beams are obtained using high NA optical devices. Thus, we chose to implement a theoretical approach based-on the angular spectrum representation of an optical field focused by an aplanatic lens, proposed by Novotny and Hecht \cite{novotny_principles_2006}. This model takes into account the filling factor that describes how much the incoming beam is expanded relative to the size of the lens. Using this expression as initial condition in FDTD is time consuming due to numerous numerical integrations compared to the LG paraxial model, but is very effective for extremely tightly focused beams (\textit{i.e.} NA~$>0.9$). 
Several CVB examples for NA ranging from 0.3 to 1.3  and generated by the two methods can be found in our Github repository  \cite{Agez_Git_2_2021}. All simulations in the following are performed taking NA~$=0.9$ while keeping $n$~=~1 for the nanoparticle environment refractive index.
Finally, note that the expressions of the magnetic $\bf{H}$ and electric $\bf{E}$ fields for the radially and the azimuthally polarized cases are related as follows:
\begin{eqnarray}
\bf{E}_{\textrm{Azi}} & = &  Z~\bf{H}_{\textrm{Rad}}\\ \label{eq:Eazi}
\bf{E}_{\textrm{Rad}} & = &  Z~\bf{H}_{\textrm{Azi}} \label{eq:Erad}
\end{eqnarray}\noindent where $Z=\sqrt{\mu/\varepsilon}$ is the wave impedance, $\mu$ and $\varepsilon$ are the magnetic permeability and the electric permittivity of the medium, respectively. Since in our numerical method the wave impedance is normalized to 1, the magnetic and the electric amplitudes can be directly compared in all the results presented hereafter.

\subsection{Silicon nanosphere excited by focused cylindrical vector beams}
\begin{figure*}
\includegraphics*[width=\textwidth]{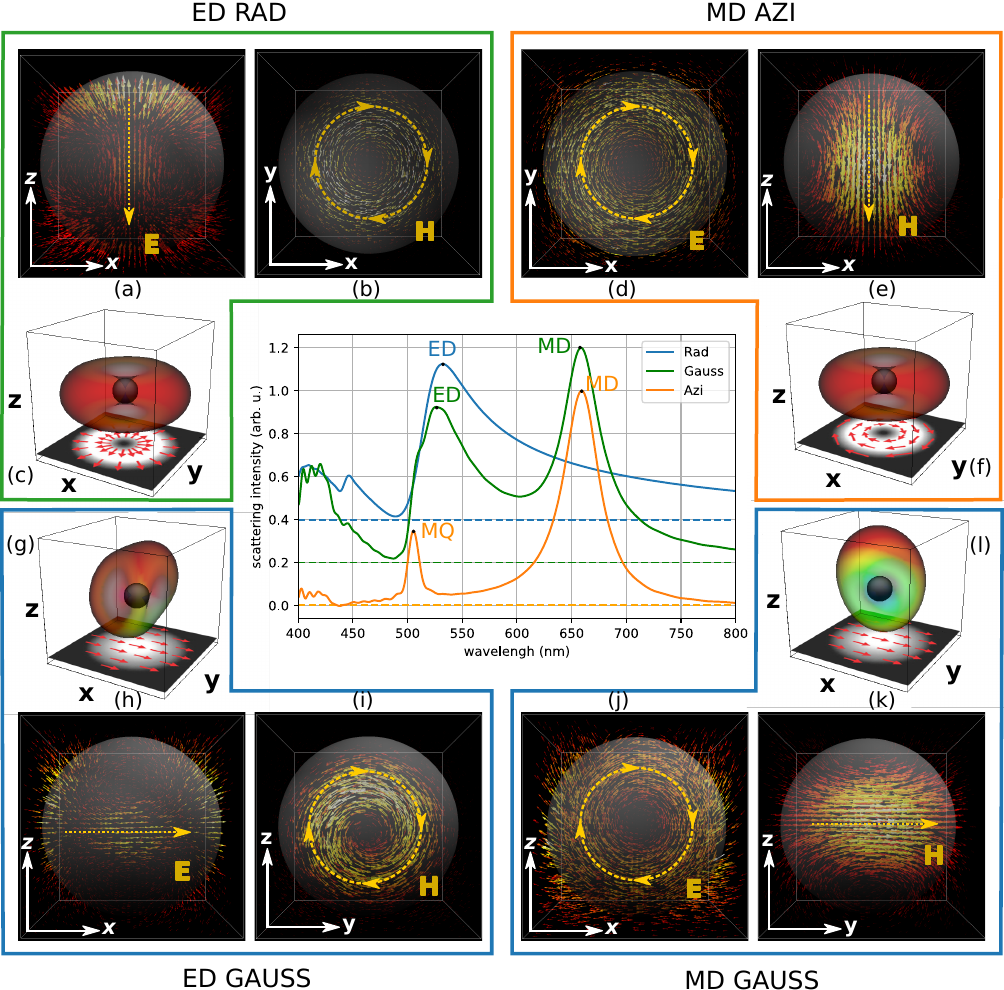}
\caption{Center panel: Scattering spectra of a 165 nm diameter Si nanosphere as function of excitation polarization. The ED (rad-CVB) and MD (azi-CVB) modes are normalized with respect to the corresponding modes of the lin-GB case. Electric ((a), (d), (h), (j)) and magnetic ((b), (e), (i), (k)) vector fields for the different resonance modes and excitation polarizations (ED with rad-CVB (green panel), MD with azi-CVB (orange panel), ED and MD with Gaussian beam (blue lower panels)); (c), (f), (g), and (l) show the far-field scattering patterns characteristic of a dipole resonance, with different dipole orientations given by the yellow arrows in (a), (e), (h) and (k). The incident light propagates from the bottom to the top; the incident beam polarization is represented in the $xy$-plane using red arrows.}
\label{fig:sphere}
\end{figure*}
Despite their simple shape and high symmetry, high-refractive index nanospheres such as Si NS exhibit more or less complex resonance spectra as function of size (radius $R$) when excited by a nonpolarized plane wave. The lowest order resonance, occurring at a wavelength in the material $\lambda/n_{Si}$ commensurate with the sphere perimeter 2$\pi$~$R$, corresponds to a magnetic dipole (MD) resulting from a curl-like electric displacement field (See Fig.~\ref{fig:sphere}). With increasing size, higher order modes may appear starting with the electric dipole (ED), then the magnetic quadrupole (MQ), etc.

Several authors \cite{wozniak_selective_2015, manna_selective_2020, das_dark_2017} have reported that using focused beams with different polarization states yields to very different resonance spectra. A linearly polarized Gaussian beam gives a spectrum that is similar to that of a linearly polarized plane wave (assuming a spot size at waist larger than the particle diameter) \cite{zambrana-puyalto_excitation_2012}. The scattering spectrum of a Si NS illuminated by an azimuthally (resp. radially) polarized focused CVB, on the the other hand, is very different as only the magnetic (resp. electric) multipolar modes are excited (See the scattering spectra in the center panel of Fig.~\ref{fig:sphere}). By changing the beam shape and polarization, which is fairly simple experimentally, it is thus possible to switch on and off different optical resonances of the nanosphere \cite{wozniak_selective_2015}.

\begin{figure*}
\includegraphics*[width=\textwidth]{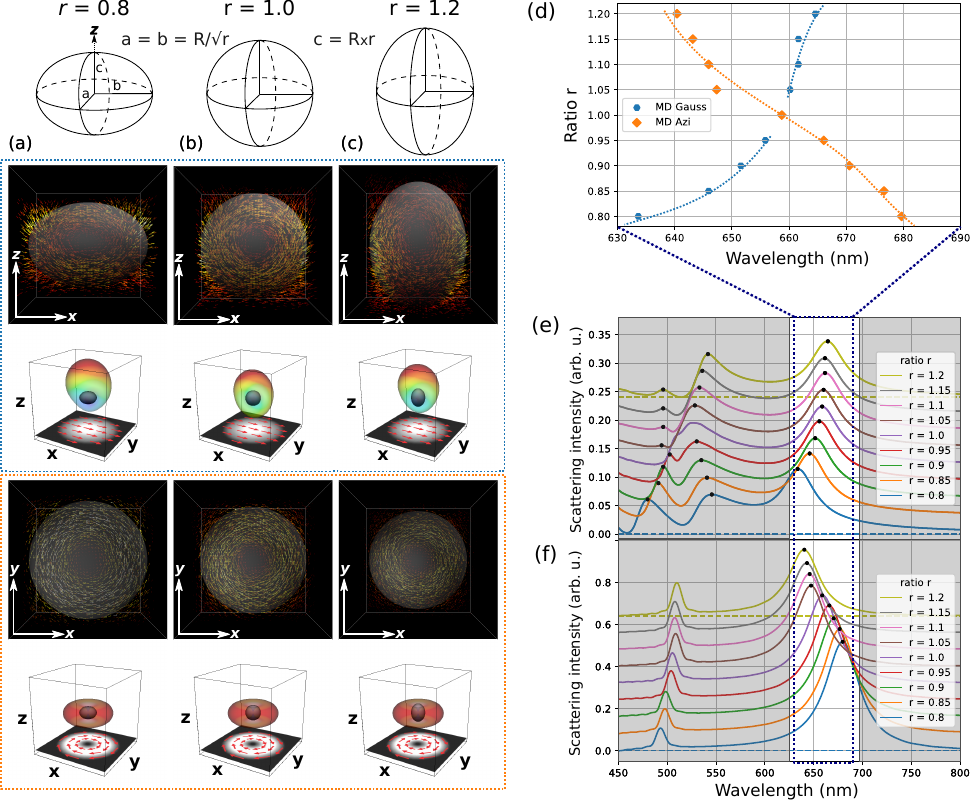}
\caption{Si nanospheroids from (a) oblate (compressed along $z$-symmetry axis) to (c) prolate (elongated along $z$-symmetry axis). The sphere has a diameter of 165 nm. Vector fields and far-field scattering patterns of the MD mode with Gaussian beam (upper blue panel) or azi-CVB (lower orange panel) excitation. The incident light propagates from the bottom to top; the beam polarization is represented in the $xy$-plane using red arrows. (e) (resp. (f)) shows the scattering spectra as function of the deformation ratio $r$ under lin-GB (resp. azi-CVB) excitation. The spectra are translated for better clarity, and the dashed horizontal line gives the zero scattering for the extreme $r$ values. (d) shows the opposite shift of the MD resonance wavelength as function of $r$ and excitation polarization.}
\label{fig:spheroid_azi}
\end{figure*}
In addition, it must be emphasized that the magnetic (resp. electric) multipolar modes excited by the azimuthally (resp. radially) polarized focused CVB are not the same as the corresponding modes observed under linearly polarized Gaussian beam excitation (lin-GB).
This is obvious for instance by comparing the MD resonance in the case of azimuthally  polarized focused CVB (azi-CVB) and lin-GB  excitations. The electric displacement current lies in different planes (Fig.~\ref{fig:sphere}(d,j)), and the resulting magnetic dipole and magnetic vector field  have a different orientation (Fig.~\ref{fig:sphere}(e,k)).
Under lin-GB excitation, the displacement current lies in the $xz$-plane (Fig.~\ref{fig:sphere}(j)) and the magnetic dipole is oriented along the $y$-axis, perpendicularly to both the propagation direction and polarization of the incident wave (Fig.~\ref{fig:sphere}(k)). On the other hand, under azi-CVB illumination, the displacement current lies in the $xy$-plane, and the magnetic dipole is parallel to the $z$-axis due to the strong magnetic field longitudinal component at the beam waist (Fig.~\ref{fig:sphere}(e-f)) \cite{novotny_principles_2006}.

The same analysis can be made for the ED resonance. With lin-GB, the electric dipole is roughly aligned with the incident polarization along the $x$-axis (Fig.~\ref{fig:sphere}(h)). For the radially polarized CVB (rad-CVB), it is found along the $z$-axis due to the to strong electric field longitudinal component at the beam waist (Fig.~\ref{fig:sphere}(a-c)) \cite{novotny_principles_2006}.
The dipole orientation as function of the excitation beam is also illustrated in Fig.~\ref{fig:sphere}(c,f,g,l), where the torus-like far-field scattering patterns correspond to electric or magnetic dipoles parallel to the $z$-axis (azi- and rad-CVB, Fig.~\ref{fig:sphere}(c,f)), or along the $x$-axis (G-beam, Fig.~\ref{fig:sphere}(g,l)). It is also interesting to note that, due to a full cylindrical symmetry around the optical axis $z$, the MD (resp. ED) radiation pattern for the azi-CVB (resp. rad-CVB) is closer to a perfect dipole than the one obtained by lin-GB.
All these points have to be taken into account when interpreting experimental and theoretical results. Hence, although the same resonance modes are found at the same wavelength in the case of highly symmetric structures such as the sphere, scattering patterns cannot be extrapolated from the lin-GB case to the other configurations. This will be also the case for the near-field maps as shown latter.

\subsection{Silicon nanospheroid  excited by focused cylindrical vector beams: influence of shape anisotropy}
The shape influence is investigated by considering oblate and prolate nanospheroids. Starting from a sphere, a deformation is applied along one of the axis (referred to as the symmetry axis) at constant volume. The spheroid shape (oblate or prolate) is defined by the aspect ratio $r$, equal to the symmetry axis half-length divided by the sphere radius $R$. $r>1$ (resp. $r<1$) corresponds to a prolate (resp. oblate) spheroid. An example of deformation along the $z$-axis is given in Fig.~\ref{fig:spheroid_azi}(a-c).
\begin{figure*}[!ht]
\includegraphics*[width=\textwidth]{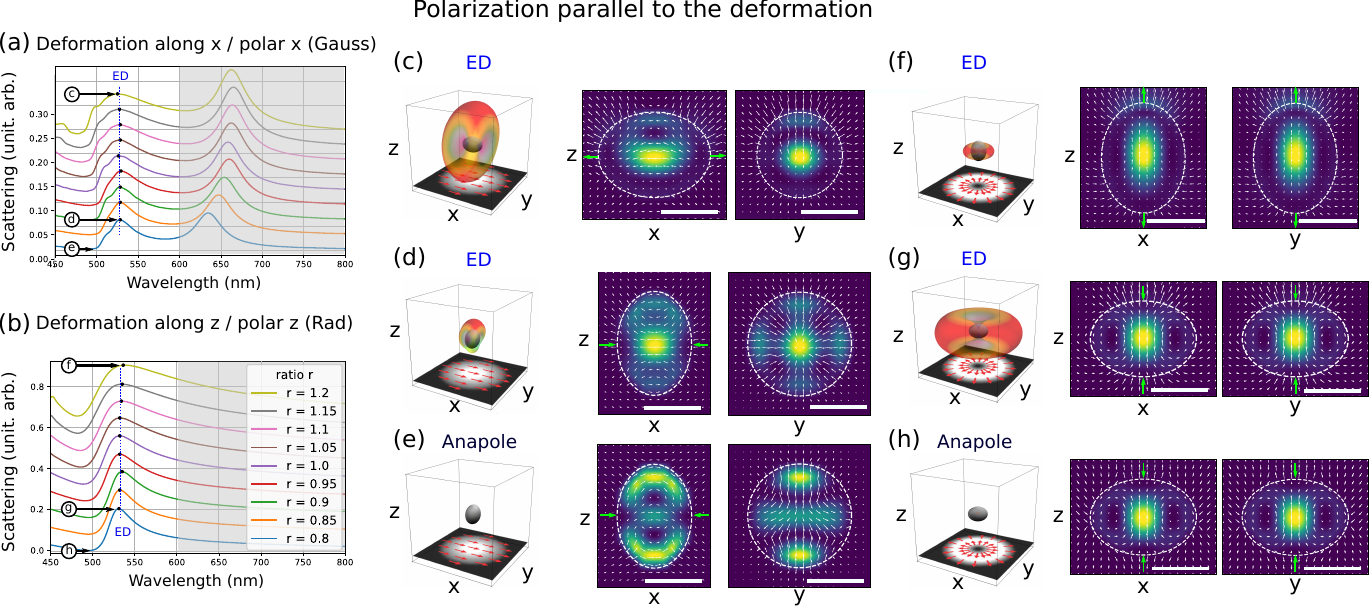}
\caption{Scattering spectra as function of the aspect ratio $r$ for polarization parallel to the symmetry axis: (a) lin-GB and $x$-symmetry axis, and (b) rad-CVB and $z$-symmetry axis.  When indicated, the dashed horizontal line gives the zero scattering. ((c)-(h)) Far-field scattering patterns and vector field maps giving to the comparison of the ED resonance and the anapole excited by either lin-GB or rad-CVB for $r$=1.2 and for $r$=0.8. The far-field patterns are normalized to the maximum value within a given kind of excitation. The intensity and vector field maps are normalized independently. Green arrows indicate the deformation axis.}
\label{fig:spheroid_rad}
\end{figure*}
\subsubsection{Magnetic dipole resonance behavior as function of excitation}
 In the case of the example of Fig.~\ref{fig:spheroid_azi}(a-c), a wavelength shift is expected for the magnetic dipole resonance when changing the excitation from lin-GB to azi-CVB, except in the case of the sphere $r=1$.
 For the azi-CVB excitation, the expected behavior  is a redshift (resp. blueshift) of the MD resonance wavelength in the case of the oblate (resp. prolate) spheroid compared to the sphere case.?
 Indeed, as seen in Fig.~\ref{fig:spheroid_azi} (lower orange frame) the disk radius enclosing the doughnut shape of the  electric displacement current in the $xy$-plane is larger (resp. smaller), than the disk radius in the sphere. At the same time, the wavelength of the MD mode excited by the $x$-axis-polarized lin-GB follows a shift in the opposite sense: blueshift in the oblate case, due to the smaller dimension in the $xz$-plane, and redshift in the prolate case, due to the larger dimension in the same plane (see Fig.~\ref{fig:spheroid_azi}, upper blue frame). The scattering spectra showing the MD mode wavelength shift as function of the aspect ratio $r$ for each polarization are given in Fig.~\ref{fig:spheroid_azi}(e-f), as well as the crossover of the curves giving the MD resonance position (Fig.~\ref{fig:spheroid_azi}(d)).

In summary, in the case of spheroids, switching between lin-GB and azi-CVB leads simultaneously to a change of the magnetic dipole orientation, hence the far-field scattering pattern, and to a shift in the opposite way of the MD resonance wavelength as function of the aspect ratio $r$.

Considering the deformation applied along the $y$-axis, the discussion is very similar to the case described above, except that the MD resonance wavelength shift between the lin-GB and the azi-CVB is reversed (See Supplemental Material, Fig.~S1).
Finally, for a deformation at constant volume along the $x$-axis, the MD resonance wavelength for the two excitation beams follows the same shift with $r$ as the electric displacement current describes the same elliptic trajectory (See Supplemental Material, Fig.~S1), but still in different planes ($xz$-plane for lin-GB and $xy$-plane for azi-CVB) leading to different far-field scattering patterns. 

\subsubsection{Electric dipole resonance behavior as function of excitation}
Intuitively, a similar behavior could be expected for the ED resonance when changing the excitation from lin-GB to rad-CVB. Taking the case previously described of a deformation along the $z$-axis (Fig.~\ref{fig:spheroid_azi}(a)-(c)), the ED resonance of the oblate spheroid should be at shorter wavelength for the rad-CVB illumination (dipole along the short $z$-axis) than for the lin-GB (dipole along the long $x$-axis). In the prolate spheroid case, an opposite behavior could be expected, with a redshift of the ED resonance wavelength for the rad-CVB illumination (dipole along the long $z$-axis) compared to the lin-GB.

\begin{figure*}[!ht]
\includegraphics*[width=\textwidth]{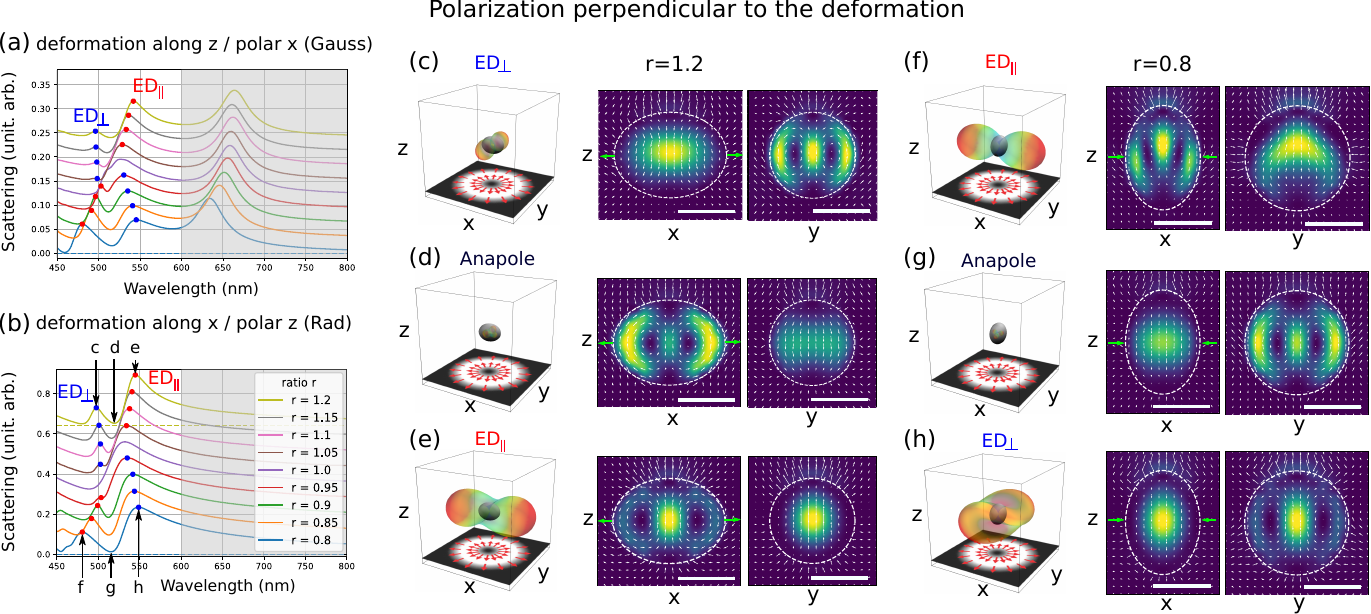}
\caption[width=\textwidth]{Scattering spectra as function of the aspect ratio $r$ for a polarization perpendicular to the symmetry axis: (a) lin-GB and $z$-symmetry axis, and (b) rad-CVB and $x$-symmetry axis.  When indicated, the dashed horizontal line gives the zero scattering. ((c)-(h)) Far-field scattering patterns and vector field maps corresponding to ED$_\perp$, ED$_\parallel$, and the anapole excited by rad-CVB for $r$=1.2 and for $r$=0.8. The far-field patterns are normalized to the maximum value. The intensity and vector field maps are normalized independently. Green arrows indicate the deformation axis.}
\label{fig:ED1ED2}
\end{figure*}

Such scenario is however not true, as illustrated in Fig.~\ref{fig:spheroid_rad}(a)-(b), where no significant ED mode wavelength shift is evidenced. while a more complex behavior occurs in the lin-GB case (Fig.~\ref{fig:spheroid_azi}(e)). In that case, and more generally when the incident light polarization is perpendicular to the spheroid symmetry axis (Fig.~\ref{fig:ED1ED2}), the ED mode is split in two peaks separate by a dip corresponding to an anapole, as discussed latter.
As for the MD mode previously described, the electric vector field and the scattering pattern at ED resonance have different orientations depending on the excitation polarization Fig.~\ref{fig:spheroid_rad}(a)-(b). 

Furthermore, the scattering pattern is distorted  in the case of the lin-GB (Fig.~\ref{fig:spheroid_rad}(c)-(d)), due to the transverse character of the polarization and the propagation direction, as well as the fact that other modes such as the MQ are simultaneously excited. Contrary, the scattering pattern in the rad-CVB case corresponds rather well to a perfect electric dipole oriented along the $z$-axis (Fig.~\ref{fig:spheroid_rad}(f)-(g)).

The zero scattering point  at the wavelength just below the ED peak (480-490 nm) (marked (e) and (h) in the scattering spectra in Fig.~\ref{fig:spheroid_rad}(a)-(b)) corresponds to an anapole, a nonradiating mode resulting from the destructive interference of electric and toroidal dipole modes (See also the far-field radiationless behavior in Fig.~\ref{fig:spheroid_rad}(e)-(h)). This toroidal dipole is a consequence of the circulating magnetic flux along a closed loop \cite{parker_excitation_2020, miroshnichenko_nonradiating_2015, wei_excitation_2016, yang_nonradiating_2019}. In highly anisotropic nanostructures, such as the spheroids considered here, the loop discontinues reducing the system to a transverse magnetic quadrupole formed by two anti-parallel magnetic moments, as reported by Miroshnichenko \textit{et al.} \cite{miroshnichenko_nonradiating_2015}. This does not prevent the formation of an anapole. Examples showing either the closed loop or the transverse MQ as function of the deformation ratio are given in Supplemental Material.  The toroidal dipole vector field direction is opposite between the ED mode and the anapole mode, resulting in a toroidal dipole in phase (ED mode) or out-of-phase (anapole mode) with the electric dipole of the ED mode.
In any case, the phase difference between the toroidal dipole (or the transverse magnetic quadrupole) and the electric dipole drives the far-field behavior (See also Appendix A, Fig.~\ref{fig:ED&TMQ}).

Compared to the previous case, when the symmetry axis is perpendicular to the excitation polarization (Fig.~\ref{fig:ED1ED2}(a)-(b)), the ED mode behavior is totally different and is split, giving a dip between two peaks, hereafter referred to as ED$_\perp$ and ED$_\parallel$. The antiresonance mode between ED$_\perp$ and ED$_\parallel$ can be attributed to an anapole, in agreement with the near-zero scattering cross-section in the spectra and the nonradiating patterns in Fig.~\ref{fig:ED1ED2}(d)-(g).
The far-field behavior of ED$_\perp$ and ED$_\parallel$ resonances is also striking, as they exhibit highly directional scattering patterns along a specific axis, very different from the isotropic scattering pattern of an oscillating dipole (Fig.~\ref{fig:ED1ED2}(c),(e),(f),(h)). The scattered intensity is along the symmetry axis in the case of ED$_\parallel$, and perpendicular to the symmetry axis in the case of ED$_\perp$. The ED$_\perp$ and ED$_\parallel$ peak wavelengths exhibit a crossover as function of the deformation ratio $r$ (the peaks are merged in the sphere case, see Fig.~\ref{fig:ED1ED2}(a)-(b)). Thus, for an $x$-symmetry axis nanospheroid excited by rad-CVB, the ED$_\parallel$, whose scattered intensity is mainly directed along the $x$-axis in the case of a deformation ratio $r=0.8$, becomes a ED$_\perp$ whose scattered intensity is mainly directed along the $y$-axis in the case of a deformation ratio $r=1.2$ (Fig.~\ref{fig:ED1ED2}(c),(f)), and reciproqually (Fig.~\ref{fig:ED1ED2}(e),(h)).
Contrary to the anapole, both the resonance and the highly directional scattering pattern at the wavelength corresponding to ED$_\perp$ and ED$_\parallel$ resonances are due to the constructive interference of the electric dipole mode and either the transverse MQ or the  toroidal dipole mode. The resonance at shorter (resp. longer) wavelength corresponds to the electric field vortices positioned symmetrically to the $z$-axis-oriented electric dipole in the smaller (resp. larger) section of the spheroid. 
The behavior will be similar in the other cases where the same behavior of ED mode split by the anapole is encountered, including the Gaussian beam linearly polarized along the $x$-axis illuminating spheroids with either the $y$- or $z$-symmetry axis (See for instance Fig.~\ref{fig:ED1ED2}a). The difference with the rad-CVB is that the scattering will not be totally cancelled at the anapole wavelength between ED$_\perp$ and ED$_\parallel$ due to the presence of the MQ mode in the same wavelength range, which is excited simultaneously with the other modes. This can be seen in Fig.~\ref{fig:ED1ED2}, where the scattering intensity drops to zero between ED$_\perp$ and ED$_\parallel$ in the rad-CVB case (Fig.~\ref{fig:ED1ED2}(b)), but lies slightly above zero in the lin-GB case (Fig.~\ref{fig:ED1ED2}(a)).

\begin{figure*}
\includegraphics{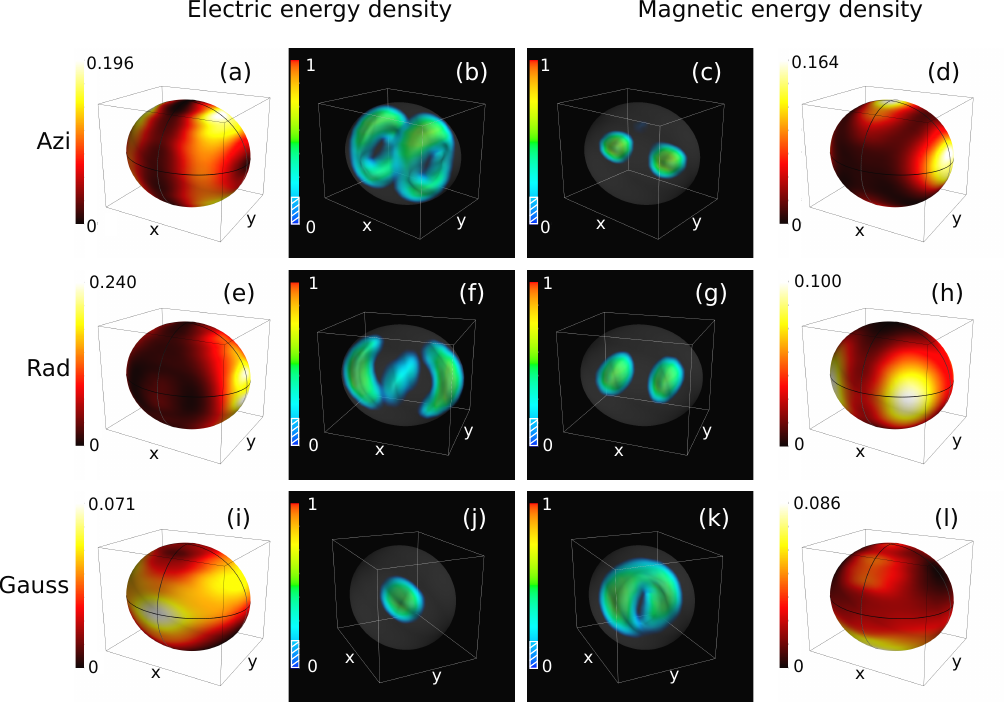}
\caption{Different near-fields maps obtained by switching the polarization (azi-CVB, rad-CVB and  $x$-axis-polarized lin-GB) of a 516 nm wavelength excitation. The Si nanostructure is an oblate spheroid with a deformation ratio of $r$=1.15 along the $x$-symmetry axis. The colored maps in the left and right columns show the electric and magnetic near-field, respectively, at 5~nm outside the spheroid surface. Images with black background depict the 3D distribution of electric and magnetic energy density inside the spheroid for each kind of excitation. They are characteristic of a MQ mode (upper row), an anapole (middle row), and a ED mode (lower row) (See also the scattering spectra in Fig.~3(b-c)). Each color bar is normalized by the local maximum energy density. For clarity, the values corresponding to the hatched part in the scale bar are not plotted.} 
\label{fig:switch}
\end{figure*}

\section{Engineering near and far field optical response of silicon nanospheroids using cylindrical focused beams} 
As shown in the previous sections, it is possible to tune the optical response of a Si sphere or spheroid by changing the polarization of the incident light. This includes triggering either electric or magnetic modes using rad-CVB or azi-CVB excitation. In addition, it is possible by choosing properly the spheroid dimensions, to tune the intensity and the directivity of the scattered light at a given wavelength.
To illustrate, we consider a spheroid with a deformation ratio $r=1.15$ along the $x$-axis illuminated by a 516 nm wavelength radiation. Fig.~\ref{fig:spheroid_rad} and Fig.~\ref{fig:switch} reveal that this wavelength corresponds to the ED resonance in the lin-GB case, to the MQ mode in the azi-CVB case, and to an anapole in the rad-CVB case. The first one is strongly radiating in the far-field while the others are considered dark or nonradiating. The spheroid should thus be invisible under CVB excitation at the 516 nm wavelength, and not under lin-GB.
Another interest in using different Laguerre-Gaussian beams is to modify the electric and magnetic near-field intensity distribution, which is an important parameter controlling the excitation of quantum emitters positioned in the vicinity of the nanoantenna \cite{bidault_dielectric_2019, poumirol_unveiling_2020, albella_low-loss_2013}.

The example of Fig.~\ref{fig:switch} shows that the maximum intensity of the electric near field has roughly the same value, but is located at different spatial positions around the spheroid. As a consequence, optimizing the nanoantenna - emitter couple needs to spatially position precisely the emitting dipoles in the nanostructure near field, as well as their orientation.
Taking the example of Transition Metal Dichalcogenide (TMD) monolayer wrapped around Si nanostructures \cite{sortino_dielectric_2020, poumirol_unveiling_2020}, the near field hot-spot leading to an increased number of excitons would be located at different positions according to the excitation polarization. Thus, in the case of the lin-GB, more excitons would be generated in the TMD monolayer on each side of the spheroid along the $x$-axis, while they would be created on each side of the spheroid along the $y$-axis in the rad-CVB case. 
In addition to this process at the excitation wavelength, the nanostructure shape could be calculated in order to make a resonance mode wavelength coincide with the exciton emission one. Furthermore, since the emitting dipoles are preferentially generated at different positions around the nanoantenna, the photoluminescence (PL) intensity pattern is expected to be very different depending on the excitation beam polarization. Note also that the PL emission directivity could be further modified by the orientation of the emitting dipoles as, depending on the kind of TMD, the emitting dipole can be parallel (neutral exciton), or perpendicular (dark exciton in specific materials such as WSe$_{2}$) to the TMD layer  \cite{poumirol_unveiling_2020}.

The spatial separation of electric and magnetic near field is important for controlling emitters that support both electric and magnetic dipolar transitions, such as rare-earth ions \cite{bidault_dielectric_2019,chacon_measuring_2020}. Selective excitation of the magnetic dipole transition can be obtained thanks to resonant azi-CVB excitation \cite{kasperczyk_excitation_2015}. By combining azi-CVB excitation and locally enhanced value of the magnetic near-field, this effect could be further improved.
In Fig.~\ref{fig:switch}, the maps of the electric and magnetic energy density at 5 nm outside the surface of the spheroid show the electric and magnetic hot spots that can be manipulated by changing the excitation beam polarization at the same wavelength. In addition, the position and the nature (electric or magnetic) of the emitting dipole exciting a resonance mode (for instance the magnetic dipole resonance, at longer wavelength than the excitation tuned to ED or MQ resonance, see Fig.~\ref{fig:sphere}) would lead to specific emission patterns \cite{bidault_dielectric_2019, sugimoto_magnetic_2021}.

\section{Cylinders with circular or elliptic sections}
\begin{figure}
\includegraphics[width=\columnwidth]{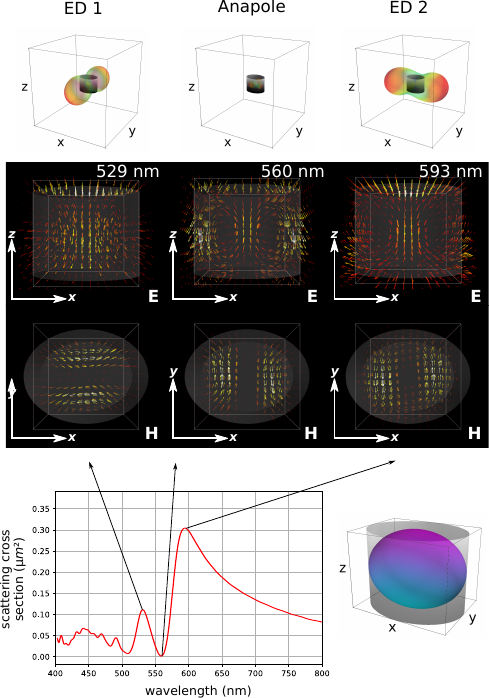}
\caption[width=\columnwidth]{ED mode split in ED$_\perp$ and ED$_\parallel$ peaks separated by the anapole mode in the case of a Si nanocylinder ($r=1.2$  with respect to the $x$-axis), illuminated by a rad-CVB propagating from bottom to top along the $z$-axis. The vector field maps show either the toroidal dipole or the transverse quadrupole mode canceling totally (anapole at 560 nm) or partially (ED$_\perp$ at 529 nm, ED$_\parallel$ at 593 nm) the electric dipole parallel to the $z$-axis. The far-field patterns are normalized showing the quasi-uniaxial strong scattering of ED$_\perp$ and ED$_\parallel$, and the nonradiating behavior of the anapole (zero scattering intensity).}
\label{fig:anapole_rad_cylinder}
\end{figure}

For applications, a top-down approach involving electron lithography and reactive ion etching is commonly used to produce Si NS from Silicon on Insulator or similar substrates \cite{baranov_all-dielectric_2017, wiecha_evolutionary_2017}. Such approach is however not very suited to obtain spheroidal NS, but instead wires with cubic or rectangular section or cylinders with circular or elliptic section.
To compare with spheres and spheroids, we take the example of cylinders with circular and elliptic sections, respectively. Starting from a nanocylinder of height $H$ and a circular section of radius $R$=$H$, a deformation at constant volume is applied along either the cylinder axis ($z$-axis) or along a diameter (defined as either the $x$- or $y$-axis). The general behavior for the cylinders is very similar to the one of the spheres and spheroids of similar dimensions, with only a slight shift of the resonances wavelengths. The observed shift is usually a redshift due to a larger volume of the cylinder compared to the spheroid (the sphere or the spheroid is circumvented in the corresponding cylinder with the circular or elliptic section, as shown in Fig.~\ref{fig:anapole_rad_cylinder}). The scattering intensity spectra for various nanocylinders and excitations are given in Supplemental Material (Fig.~S3).
The chosen example of Fig.~\ref{fig:anapole_rad_cylinder} corresponds to a cylinder with a deformation ratio $r=1.2$ along the $x$-axis illuminated by a radially polarized beam. The elliptic section dimensions are thus $R.r$ (long axis), and $R/\sqrt{r}$ (short axis), the height being $H/\sqrt{r}$. This configuration can be compared to the sphere of radius $R=165$~nm and the prolate spheroid shown in  Fig.~\ref{fig:ED1ED2}(c)-(e). The electric and magnetic vector fields are very similar (including the phase change between the electric dipole and the transverse magnetic quadrupole or the toroidal dipole, when looking at ED$_\perp$, ED$_\parallel$, and the anapole), as well as the patterns of the far field scattered intensity.

\section{Conclusion}

To summarize, we provide a Python toolkit that describes the various CVB sources and allows FDTD simulations to be performed using the Meep open-source software. We demonstrate that, by properly selecting the shape of a silicon nanospheroid and the polarization of the incident cylindrical vector beam, the optical response of the nanoantenna can be adjusted, either in the near-field or in the far-field. An example of selective excitation at a fixed wavelength of an electric dipole, magnetic quadrupole, or anapole mode as a function of CVB polarization is described (linear, azimuthal, radial).  At last, as spheres and spheroids are certainly not the best fabrication-friendly candidates for large scale reproducible production of nanostructures, we show that the trends observed in nanospheroids are also found in nanostructures obtained by techniques based on electron beam lithography, such as nanocylinders with circular or elliptic section.

\begin{acknowledgments}
The authors acknowledge funding from Agence Nationale de la Recherche under projects HiLight (ANR-19-CE24-0020-01) and EUR NanoX 2DLight (ANR-17-EURE-0009) in the framework of the Programme des Investissements d'Avenir, and CALMIP for the access to the supercomputing facility Olympe under project P19042.
\end{acknowledgments}

\appendix

\section{Electric dipole and tranverse magnetic quadrupole coupling}

\begin{figure}[ht!]
\includegraphics*{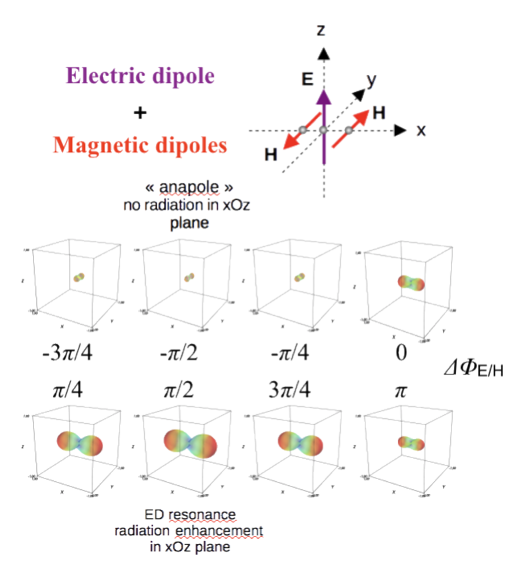}
\caption[width=\columnwidth]{Scheme showing the electric dipole and the two magnetic dipoles forming the transverse magnetic quadrupole, and far field scattering patterns of the coupled ED and transverse MQ as function of phase difference $\Delta\Phi_{E/H}$ between $E$ and $H$.}
\label{fig:ED&TMQ}
\end{figure}

 Fig.~\ref{fig:ED&TMQ} shows the coupling of an electric dipole and a transverse magnetic quadrupole as function of the phase difference between the electric and magnetic fields radiated by the electric and magnetic dipoles, respectively.
 The -$\pi$/2 phase difference corresponds to the anapole mode and the +$\pi$/2 phase difference corresponds to an ED-like resonance mode, illustrating the nonradiating behavior of the anapole and the high directivity of the ED$_\perp$ and ED$_\parallel$ scattered intensity.
 
 The example in Fig.~\ref{fig:ED&TMQ} is analogous to the ED$_\parallel$ peak at 480 nm shown in Fig.~\ref{fig:ED1ED2}(e). Aligning the magnetic dipoles along the $x$-axis also rotates the scattering pattern  by 90$^{\circ}$ to give a system analogous to the ED$_\perp$ peak at 497 nm in Fig.~\ref{fig:ED1ED2}(c).

\nocite{*}
\bibliography{Montagnac_CVB_spheroid_main.bib}
\end{document}